%% file: skeleton.tex
\documentclass[a4paper,11pt]{article}
\usepackage{pos}
\usepackage[export]{adjustbox}
\usepackage{mciteplus}
\usepackage{amsmath}

\usepackage{ifthen}
\newboolean{articletitles}
\setboolean{articletitles}{true}

\usepackage{lineno}
\title{Time-integrated CP asymmetries in meson and baryon decays}

\author*[a,b]{Alex Gilman}

\onbehalf{for the the LHCb and Belle II Collaborations}

\affiliation[a]{Department of Physics, University of Cincinnati,\\
  Cincinnati, Ohio 45221, USA}

\affiliation[b]{Department of Physics, University of Oxford,\\
Oxford OX1 3PU, UK\\
Presented at the 32nd International Symposium on Lepton Photon Interactions at High Energies, Madison, Wisconsin}

\emailAdd{alexander.leon.gilman@cern.ch}

\include{definitions}

\abstract{Measurements of CP asymmetries in hadron decays integrated over time provide access to direct CP violation, which arises from differences between the amplitudes of CP-conjugate decay processes. These proceedings present an overview of recent measurements from the LHCb and Belle II experiments, including progress on the determination of the CKM angle $\gamma$, the first observation of CP violation in baryon decays, and new studies of direct CP violation in $D$ mesons. The outlook for measurements with datasets whose collection is underway is also discussed.}

\FullConference{%
  Lepton-Photon 2025\\
  Aug. 25- Aug. 29 2025\\
  Madison, Wisconsin, USA
}

\begin{document}
\maketitle

Measurements of charge-parity (CP) violation have played a foundational role in understanding the Standard Model (SM) of particle physics. To date, CP violation has only been observed in flavor-changing interactions of quarks mediated by the weak force. Within the SM, observed CP violation is constrained through the unitarity of the Cabibbo-Kobayashi-Maskawa (CKM) matrix, which connects the mass eigenstates and weak eigenstates of quarks. These constraints can be tested through a system of experimental measurements, allowing for searches on constraints on a wide variety of beyond-Standard-Model physics scenarios\cite{Charles:2013aka,Charles:2020dfl}.

Time-integrated measurements are primarily sensitive to CP violation in decay amplitudes, where a hadron $M$ and its corresponding antiparticle $\overline{M}$ decay to a given final state $f$ (or $\overline f$, for the $\overline{M}$) at different rates. This phenomenon, known as direct CP violation, requires the interference of at least two $M \to f$ amplitudes $a_1,a_2$ with relative CP-odd weak phase $\phi$ and CP-even strong phase $\delta$. Direct CP asymmetry will then manifest according to
\begin{equation}\label{eq:ACP}
A_\text{CP} \equiv \frac{\Gamma(M \to f) - \Gamma(\bar{M} \to \bar{f})}{\Gamma(M \to f) + \Gamma(\bar{M} \to \bar{f})} \propto \frac{|a_2|}{|a_1|} \sin\phi \sin\delta.
\end{equation}

CP violation was first established through the observation of $K_L^0 \to \pi\pi$ decays in 1964 by Cronin and Fitch\cite{Christenson:1964fg}, with the first establishment of direct CP-violation in 1999 by demonstrating that the ratio $\frac{\Gamma(K_{L}^0 \to \pi^0\pi^0)}{\Gamma(K_{S}^0 \to \pi^0\pi^0)}/\frac{\Gamma(K_{L}^0 \to \pi^+\pi^-)}{\Gamma(K_{S}^0 \to \pi^+\pi^-)}$ deviates from unity through measurements by the KTeV\cite{KTeV}, NA48\cite{NA48}, E731\cite{E731},  and NA31\cite{NA31} collaborations. The first observation of direct CP violation in a heavy quark system came from measurements by the BaBar and Belle collaborations of $B^0 \to K^+\pi^-$ in 2004 \cite{BaBar:2004gyj,Belle:2004nch}. More recently, the LHCb collaboration established direct CP violation in charm decays\cite{LHCb:2019hro} through measurements of two-body charm decays in 2019. These proceedings discuss another major milestone on this topic: the first observation of CP violation in the decays of baryons\cite{LHCb:2025ray}.

\begin{figure}[hbtp]
\centering
 \includegraphics[width=0.45\textwidth]{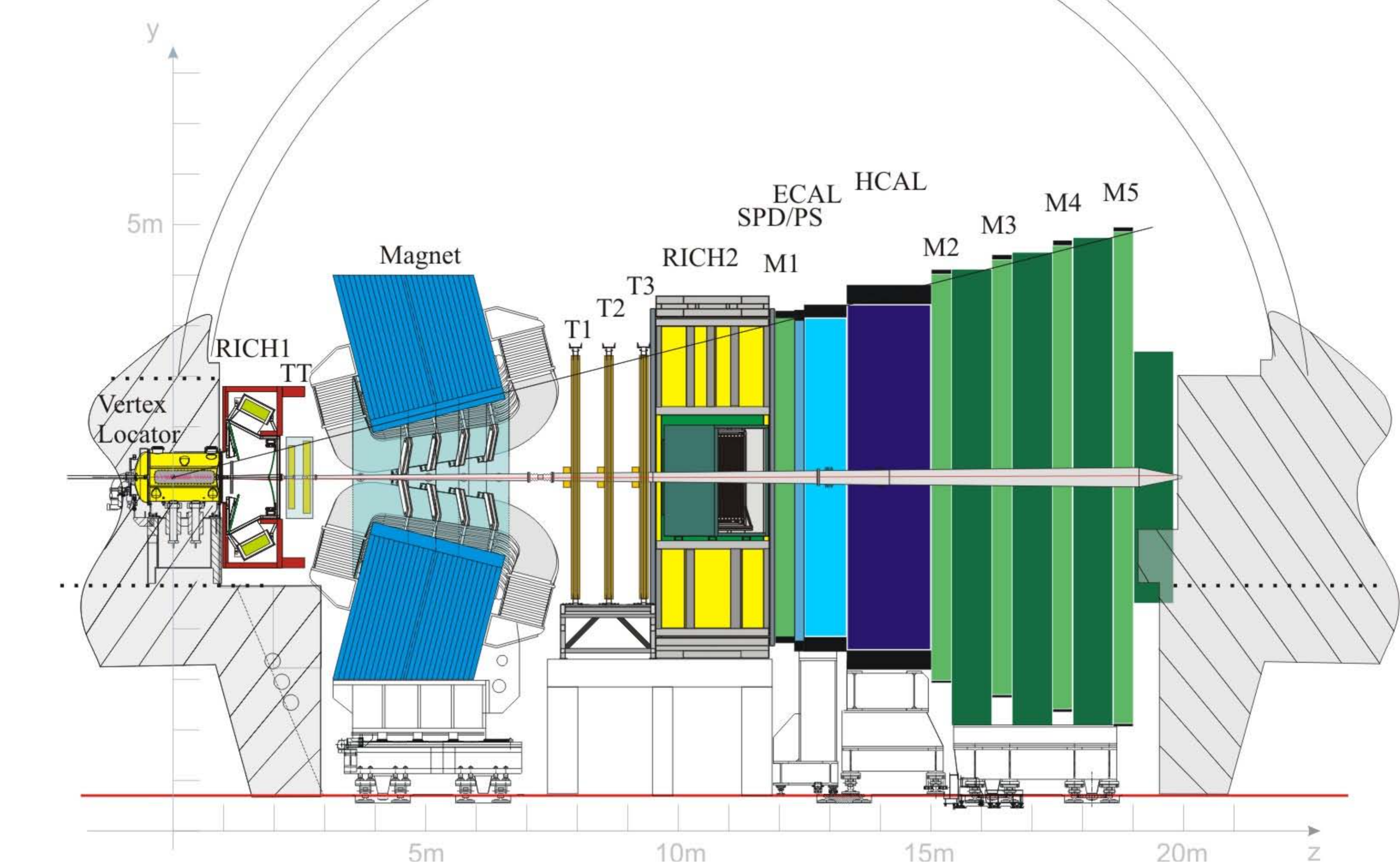}
 \includegraphics[width=0.45\textwidth]{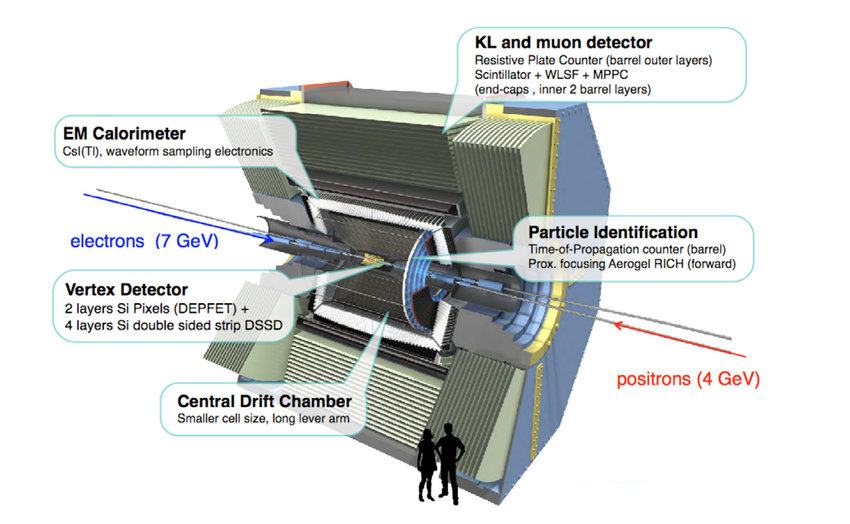}
\caption{Cross-sections of the (left) LHCb and (right) Belle II experimental setups.}
\label{fig:detectors}
\end{figure}

These proceedings focus on recent results from the LHCb and Belle II experiments. LHCb is a single-arm spectrometer at the Large Hadron Collider, operating  proton-proton collisions at center-of-mass energies of 7--13.6~TeV \cite{LHCb:2008vvz}. Following detector upgrades \cite{LHCb:2023hlw} in 2022, including the move to an all-software trigger, LHCb has collected approximately 16~fb$^{-1}$ of data during Run 3 as of Aug. 28th 2025. Belle II is a cylindrical spectrometer at the SuperKEKB $e^+e^-$ collider \cite{Belle-II:2010dht}, which operates at and above the $B\bar{B}$ threshold. Run 1 of Belle II (2019--2022) collected 0.43~ab$^{-1}$, while Run 2 has collected 0.15~ab$^{-1}$ as of as of Aug. 28th 2025. Belle II features improved acceptance, vertex resolution, and neutral particle reconstruction compared to its predecessor Belle \cite{Belle:2000cnh}, and software infrastructure that allows simultaneous analysis of Belle and Belle II data. The experimental setups of both experiments are shown in Fig.~\ref{fig:detectors}.

\section{Progress on the CKM angle $\gamma$}

The CKM angle $\gamma \equiv \arg(V_{ud}V_{ub}^*/V_{cd}V_{cb}^*)$ can be determined from the interference of tree-level amplitudes in $B \to Dh$ decays, where $h$ represents a light hadron such as $K$ or $\pi$, and the absence of a superscript on the $D$ indicates a superposition of a $D^0$ and $\overline{D}^0$ state. This determination has negligible theoretical uncertainty within the SM\cite{Brod:2013sga}, provided that both $B$ and $D$ decay amplitude parameters can be determined. The $D$ decay parameters are measured in quantum-correlated $e^+e^- \to D\bar{D}$ data from BESIII and CLEO, enabling the determination of $\gamma$ and the $B$ decay parameter space. As such, measurements of $\gamma$ provide a benchmark for CP-violation in the Standard Model, neglecting new physics at tree-level\cite{Brod:2014bfa}. The primary sensitivity comes from $B^\pm \to DK^\pm$ decays, with complementary contributions from other $B \to Dh$ modes.

\begin{figure}[hbtp]
\centering
 \includegraphics[width=0.55\textwidth,valign=c]{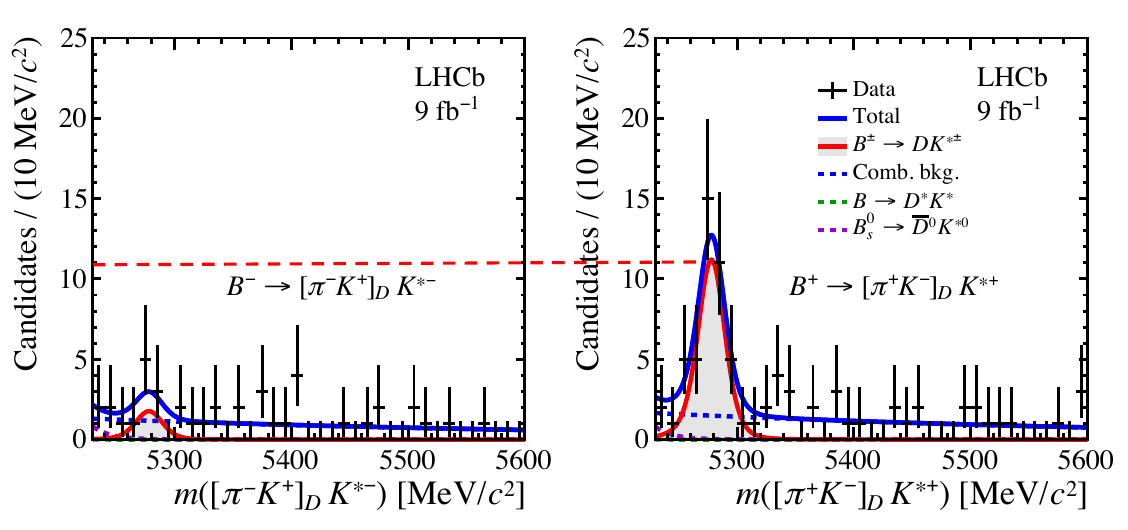}
 \includegraphics[width=0.35\textwidth,valign=c]{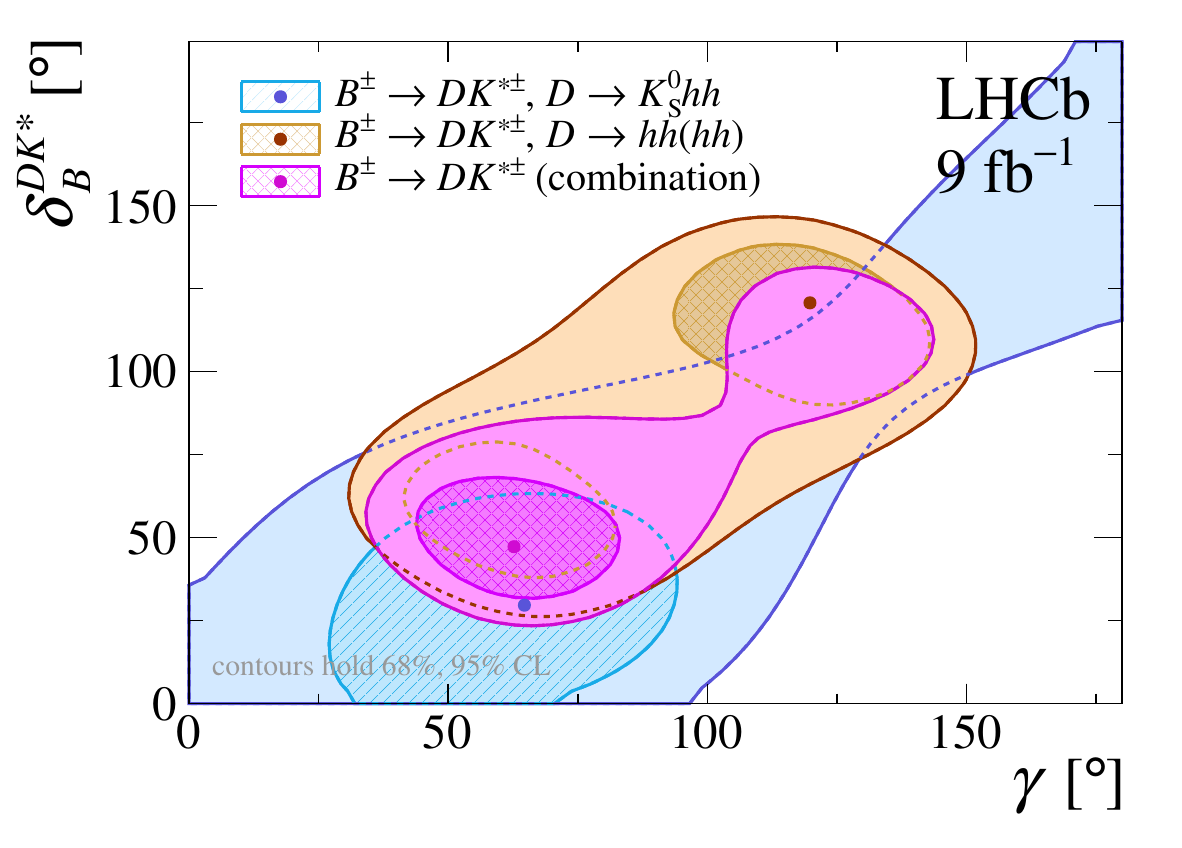}
\caption{Invariant mass distributions for $B^- \to D[\pi^-K^+]K^{*-}$ (left) and $B^+ \to D[\pi^+K^-]K^{*+}$ (center) candidates with an eye-guide to highlight the observed asymmetry. (Right) Confidence level contours in the $\gamma$--$\delta_B^{DK^*}$ plane (right) from a combined analysis of all examined $D$ decay modes of the $B^\pm\to DK^{*\pm}$ process in Ref.~\cite{LHCb:2024ett} .}
\label{fig:dkstar}
\end{figure}

A recent measurement using LHC Run 1 and Run 2 data analyzed $B^\pm \to DK^{*\pm}$ decays \cite{LHCb:2024ett}, where the $K^*(892)^\pm$ state is identified in the $K_S^0\pi^\pm$ final state. The analysis includes nine $D$ decay channels: $h'^{\pm}h^{\mp}(\pi^+\pi^-)$ and $K_S^0 h^+h^-$, where all combinations of $h',h=K,\pi$ are included. The $D$-decay amplitude parameters are constrained by measurements from the BESIII and CLEO collaborations\cite{CLEO:2010iul,BESIII:2020hpo,BESIII:2020khq,BESIII:2021eud,BESIII:2022wqs}. A combined analysis of the observed yields in all $D$ decay modes from the obtained $\gamma = (63 \pm 13)^\circ$. Examples of the fit results and the constraints on $\gamma$ are shown in Fig.~\ref{fig:dkstar}.

\begin{figure}[h]
\centering
 \includegraphics[height=0.25\textwidth,valign=c]{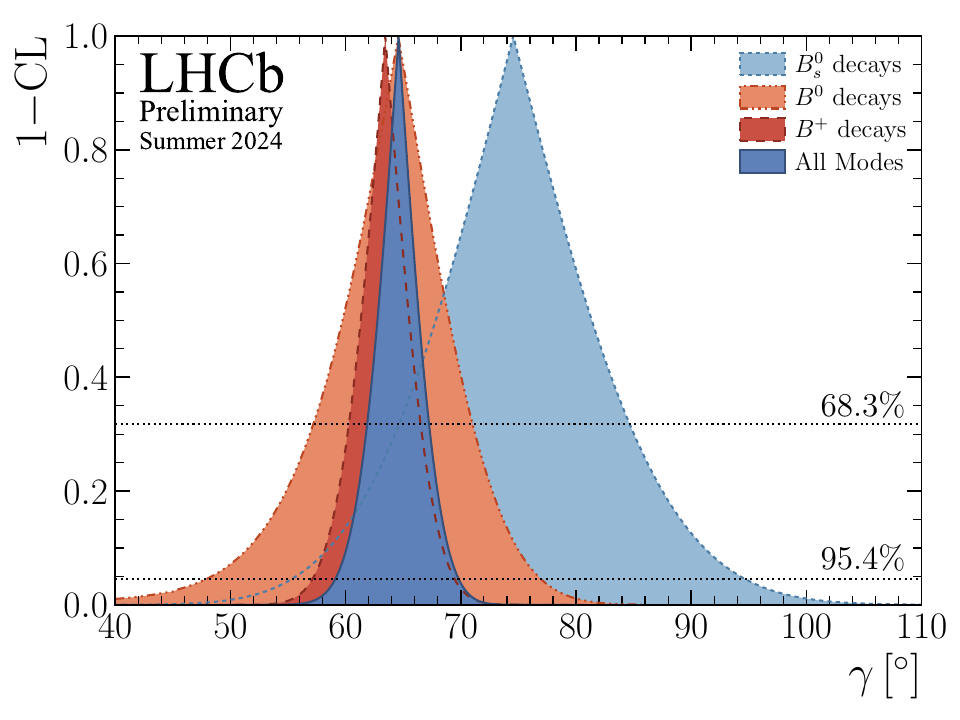}
 \includegraphics[height=0.25\textwidth,valign=c]{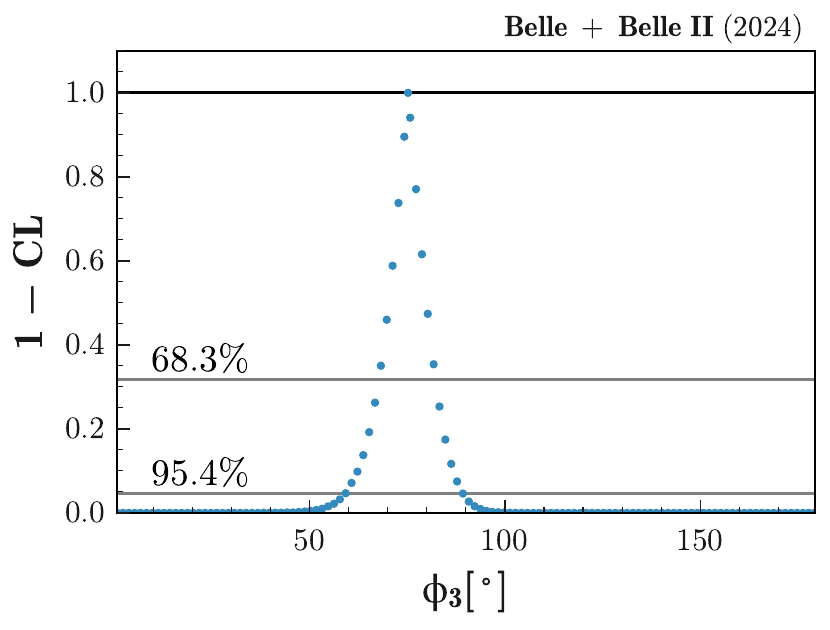}
 \includegraphics[height=0.3\textwidth,valign=c]{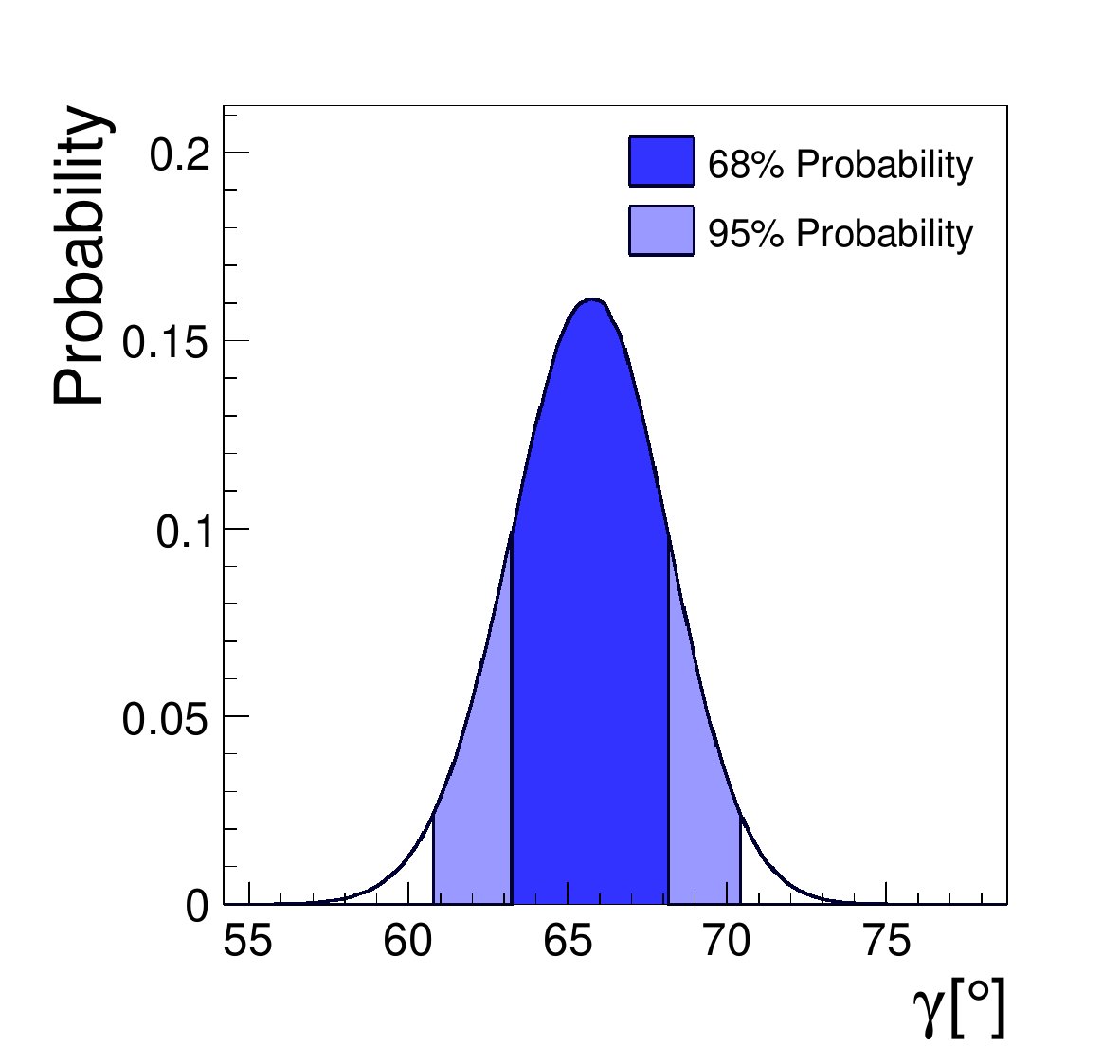}
\caption{Confidence level distributions for $\gamma$ from (left) the LHCb 2024 combination~\cite{LHCb:2024yxi}, (center) Belle/Belle II 2024 combination ~\cite{Belle:2024knt}, and (right) 2025 global analysis ~\cite{Betti:2024ldy}.}
\label{fig:gamma_combinations}
\end{figure}

The precision on $\gamma$ has improved dramatically over the past decade through the combination of many complementary measurements. These cover a variety of different $D$-decay channels, and measurements in different $B^+$, $B^0$, and $B_s^0$ decay channels. The LHCb collaboration performed a 2024 combination of all LHCb measurements, with external constraints on $D$ amplitude parameters, to find. $\gamma = (64.6 \pm 2.8)^\circ$~\cite{LHCb:2024yxi}. Similarly, a Belle and Belle II 2024 combination\footnote{The Belle and Belle II collaborations use the notation $\phi_3$ to refer to the same CKM angle which $\gamma$ refers to in these proceedings.} found $\gamma = (75.2 \pm 7.6)^\circ$~\cite{Belle:2024knt}. A 2025 global analysis incorporating over 100 observables from dozens of measurements across BaBar, Belle, Belle II, BESIII, CLEO, and LHCb experiments determined $\gamma = (65.7 \pm 2.5)^\circ$~\cite{Betti:2024ldy}. The determined confidence levels from each combination are shown in Fig.~\ref{fig:gamma_combinations}.

The evolution of the $\gamma$ measurement precision over time demonstrates the steady improvement from accumulated statistics and refined analysis techniques. Between 2015 and 2025, the precision has improved from  $\sim 7^\circ$ \cite{Auge:2014tna} to the current precision of $\pm 2.5^\circ$.

\section{Baryon decays}

While CP violation has been well-established in meson systems, its manifestation in baryon decays has eluded observation until recently. As Eq.~\ref{eq:ACP} indicates, the manifestation of CP-violation in hadronic systems within the Standard Model depends not only on the the CP-violating weak contribution to the decay amplitudes, but also on the CP-conserving contributions to decay amplitudes, largely due to strong interactions. As such, measurements of CP-violation in baryonic decays provide insight not only into the weak decays of the heavy quarks, but also on strong dynamics of baryonic systems.

\begin{figure}[h]
\centering
 \includegraphics[width=0.6\textwidth]{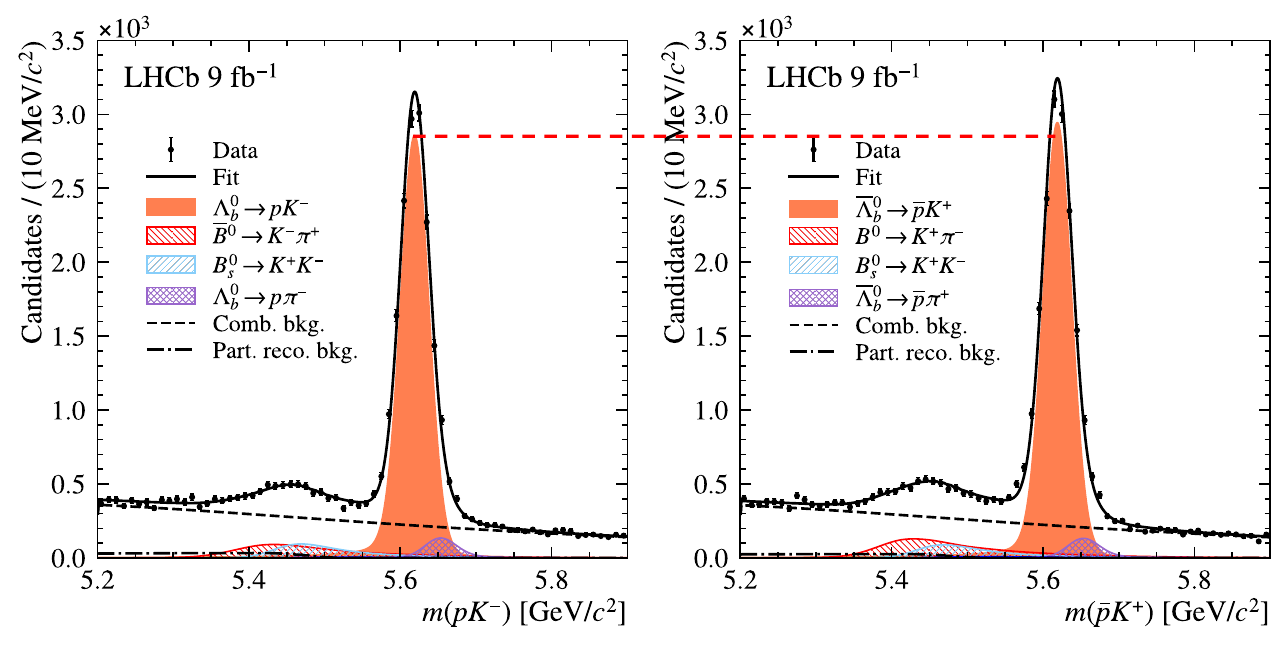}
\caption{Invariant mass distributions for (left) $\Lambda_b^0 \to pK^-$  and (right) $\bar{\Lambda}_b^0 \to \bar{p}K^+$  candidates from Ref.~\cite{LHCb:2024iis}. An eye-guide is included for ease of comparison.}
\label{fig:lambda_ph}
\end{figure}

A 2025 LHCb analysis of $\Lambda_b^0 \to pK^-$ and $\Lambda_b^0 \to p\pi^-$ decays using LHC Run 1 and Run 2 data searched for CP violation~\cite{LHCb:2024iis}. The $\Lambda_b^0 \to pK^-$ decay proceeds through the same weak transition as the well-studied $B^0 \to K^-\pi^+$ decay, which exhibits CP asymmetries of order 10\%. However, the measurement found $A_{CP}^{pK^-} = (-1.4 \pm 0.7 \pm 0.4)\%$, consistent with zero. The fit to determine the observed yields is shown in Fig.~\ref{fig:lambda_ph}. The lack of observation of CP violation in this decay channel highlights the difference in the strong dynamics between the meson and baryon systems.

A 2025 LHCb analysis with the Run 1 and Run 2 LHCb sample examined CP asymmetries in the multibody  $\Lambda_b^0 \to \Lambda h^+h'^-$ and $\Xi_b^0 \to \Lambda h^+h'^-$ decays, where $h, h' = K, \pi$~\cite{LHCb:2024yzj}. A control mode $\Lambda_b^0 \to \Lambda_c^+[\Lambda\pi^+]\pi^-$ is employed to correct for for production and detection asymmetries. The analysis found $A_{CP}(\Lambda_b^0 \to \Lambda K^+K^-) = (8.3 \pm 2.3 \pm 1.6)\%$ at $3.1\sigma$ significance, with no evidence in other channels.

\begin{figure}[h]
\centering
 \includegraphics[width=0.3\textwidth,valign=c]{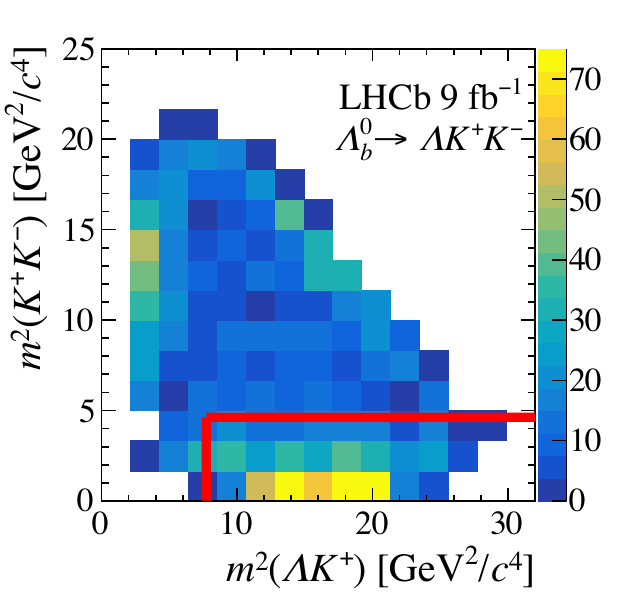} \includegraphics[width=0.65\textwidth,valign=c]{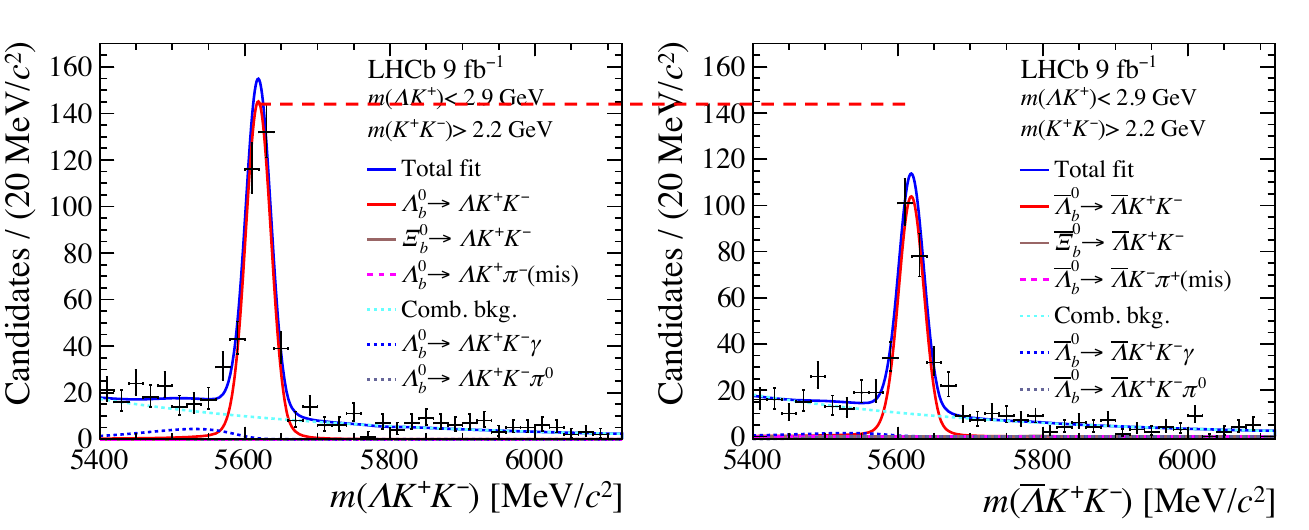}

\caption{ (Left) Region of $\Lambda_b^0 \to \Lambda K^+K^-$ phase space where largest asymmetries are observed, and (right): invariant mas distributions of selected (center) $\Lambda_b^0 \to \Lambda K^+K^-$ and (right) $\overline{\Lambda}_b^0 \to \overline{\Lambda} K^+K^-$ events in this region from Ref.~\cite{LHCb:2024yzj}, with an eye-guide for reference.}
\label{fig:lambda_kk}
\end{figure}

As the strong-phase varies across the decay phase space, asymmetries were studied in isolated regions of the phase space corresponding to resonant regions. In the resonant $\Lambda K$ region (defined by $m(K^+K^-) > 2.2$~GeV and $m(\Lambda K^+) < 2.9$~GeV), a significantly larger asymmetry was observed: $A_{CP}(\Lambda_b^0 \to R(\Lambda K^+)K^-) = (16.5 \pm 4.8 \pm 1.7)\%$ at $3.2\sigma$ significance, as shown in Fig.~\ref{fig:lambda_kk}.

The first observation of CP violation in baryon decays came from an analysis of $\Lambda_b^0 \to pK^-\pi^+\pi^-$ using LHC Run 1 and Run 2 data~\cite{LHCb:2025ray}. The control mode $\Lambda_b^0 \to \Lambda_c^+[pK^-\pi^+]\pi^-$ was employed to handle production and detection asymmetries. The inclusive asymmetry was measured to be $A_{CP}(\Lambda_b^0 \to pK^-\pi^+\pi^-) = (2.45 \pm 0.46 \pm 0.10)\%$, constituting a $5.2\sigma$ observation.

Four resonant regions were isolated for detailed study. The region containing resonances in the $p\pi^+\pi^-$ system (where $m_{pK^-} < 2.2$~GeV and $m_{\pi^+\pi^-} < 1.1$~GeV) showed the largest asymmetry: $A_{CP}(\Lambda_b^0 \to R(p\pi^+\pi^-)K^-) = (5.4 \pm 0.9 \pm 0.1)\%$, corresponding to a $6.0\sigma$ significance. Other regions showed varying levels of asymmetry, demonstrating the crucial role of resonant structures in enhancing CP violation.

\begin{figure}[h]
\centering
 \includegraphics[width=0.3\textwidth,valign=c]{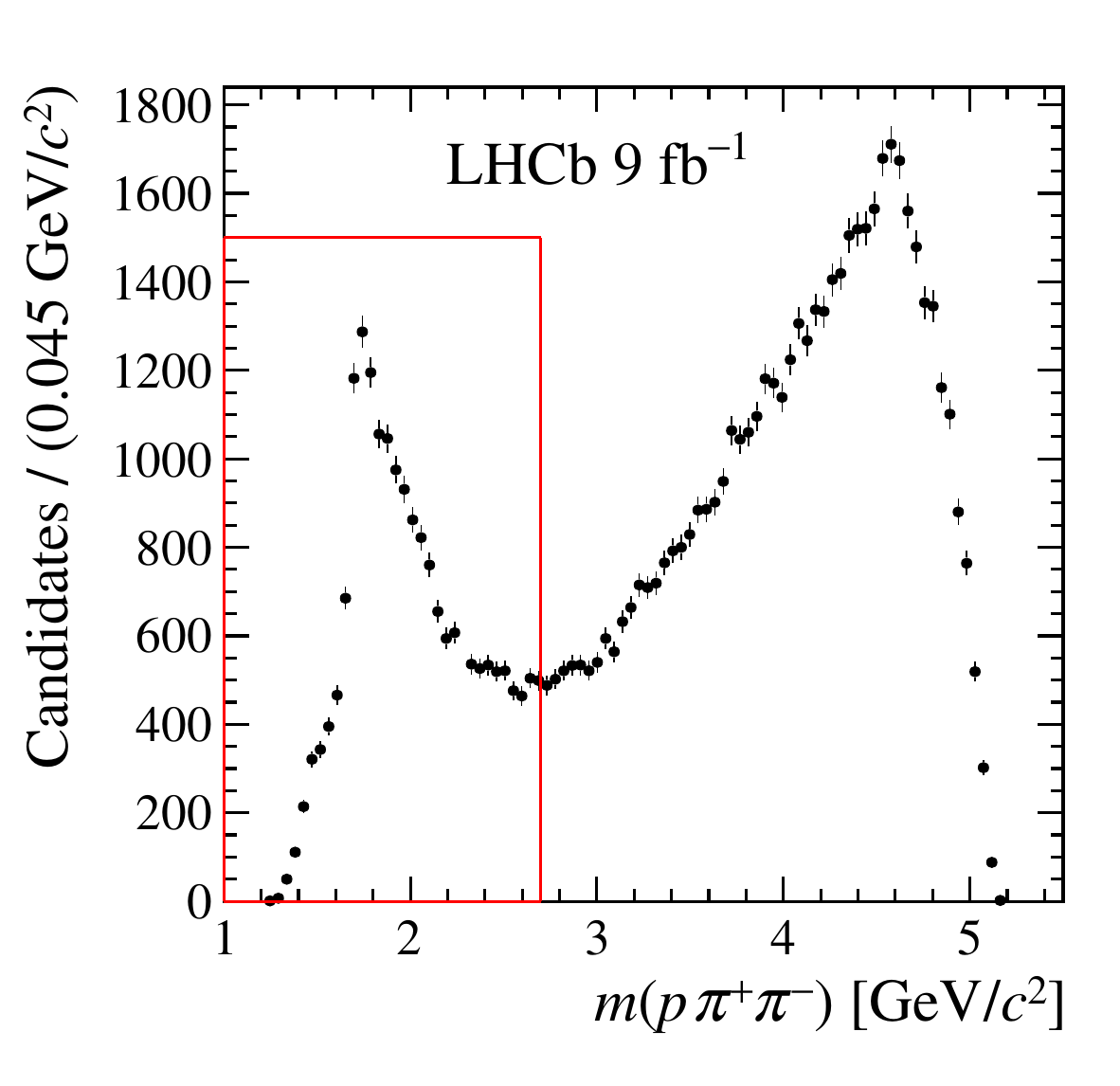} \includegraphics[width=0.65\textwidth,valign=c]{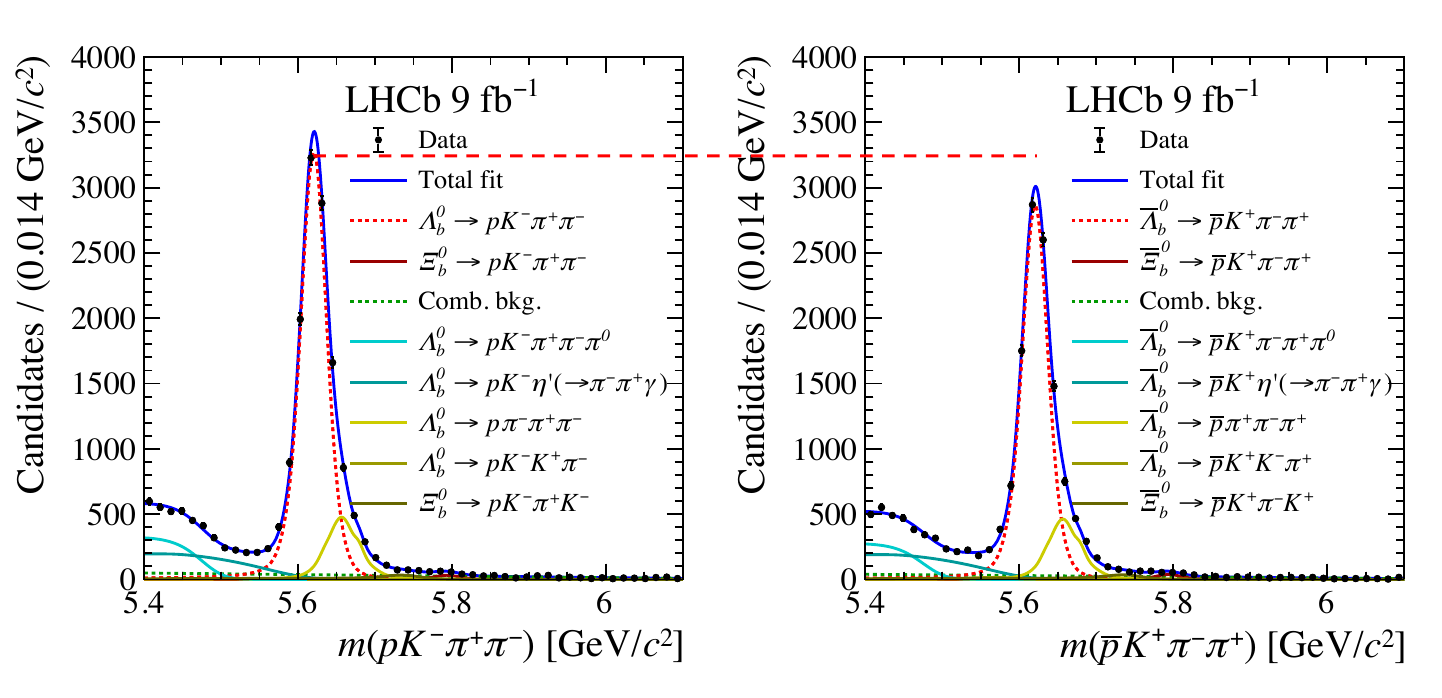}
\caption{(Left) Region of $\Lambda_b^0\to p K^-\pi^+\pi^-$ phase space where the largest asymmetry is observed, and (right): invariant mas distributions of selected (center) $\Lambda_b^0 \to pK^-\pi^+\pi^-$ and (right) $\overline{\Lambda}_b^0 \to \overline{p}K^+\pi^+\pi^-$ events in this region from Ref.~\cite{LHCb:2025ray} , with an eye-guide for reference.}
\label{fig:lambda_pkpipi_table}
\end{figure}

\begin{table}[h]
\centering
\caption{Summary of regions of the examined regions of the $\Lambda_b^0\to p K^-\pi^+\pi^-$ phase space that are examined, and the measured CP asymmetry in each region from Ref.~\cite{LHCb:2025ray}.  }
\begin{tabular}{ccc}
\hline
Decay topology & Mass region (GeV/$c^2$) & $A_{\text{CP}}$ \\
\hline
& $m_{pK^-} < 2.2$ & \\
$\Lambda_b^0 \to R(pK^-)R(\pi^+\pi^-)$ & $m_{\pi^+\pi^-} < 1.1$ & $(5.3 \pm 1.3 \pm 0.2)\%$ \\[1em]
& $m_{p\pi^-} < 1.7$ & \\
$\Lambda_b^0 \to R(p\pi^-)R(K^-\pi^+)$ & $0.8 < m_{\pi^+K^-} < 1.0$ & $(2.7 \pm 0.8 \pm 0.1)\%$ \\
& or $1.1 < m_{\pi^+K^-} < 1.6$ & \\[1em]
$\Lambda_b^0 \to R(p\pi^+\pi^-)K^-$ & $m_{p\pi^+\pi^-} < 2.7$ & $(5.4 \pm 0.9 \pm 0.1)\%$ \\[1em]
$\Lambda_b^0 \to R(K^-\pi^+\pi^-)p$ & $m_{K^-\pi^+\pi^-} < 2.0$ & $(2.0 \pm 1.2 \pm 0.3)\%$ \\
\hline
\end{tabular}
\end{table}

These results give insight into the interplay of weak and strong interactions in CP violation. The absence of observable CP violation in $\Lambda_b^0 \to ph^-$ decays at the percent level, compared to the 10--30\% effects seen in analogous two-body $B$ meson decays, suggests that strong interaction dynamics may suppress CP violation in simple baryon decay topologies. However, the observation of significant CP violation in the complex resonance structures of $\Lambda_b^0 \to \Lambda K^+K^-$ and $\Lambda_b^0 \to pK^-\pi^+\pi^-$ indicates that multibody final states with interfering resonances provide favorable conditions for CP-violating effects to manifest.

These observations establish CP violation in baryon decays as a testbed for both weak and strong interactions, and resulted in part from theoretical analysis suggesting the examined multibody decay channels as candidates to exhibit enhanced CP violation \cite{Wang:2024oyi,Han:2024kgz}.  Further experimental and theoretical analysis of baryonic decays hold great prospects for improved understanding of baryon dynamics and their impact on CP violation.

\section{D Meson Decays}

The charm sector provides a unique laboratory for studying CP violation, as the expected effects in the SM are small, making the system sensitive to potential new physics contributions. In 2019, LHCb reported the first observation of CP violation in charm decays, measuring a non-zero difference of CP-asymmetries between the $D^0\to K^+K^-$ and $D^0\to 
\pi^+\pi^-$ decays~\cite{LHCb:2019hro}. While the existence of CP violation in the charm-quark sector at the observed $\mathcal O(10^{-3})$ level is not excluded in the SM, uncertainties in theoretical calculations do not allow for a quantitative comparison\cite{Lenz:2020awd,Pajero:2022vev,Petrov:2024ujw}. Subsequently, in 2023, LHCb performed a measurement to determine direct CP asymmetries in the two channels, finding $a_\text{dir}^{K^+K^-} = (0.8 \pm 0.6) \times 10^{-3}$ and $a_\text{dir}^{\pi^+\pi^-} = (2.3 \pm 0.6) \times 10^{-3}$~\cite{LHCb:2022lry}. These measurements show approximately $2\sigma$ tension with expectations based on $d \leftrightarrow s$ exchange symmetry-breaking effects~\cite{Pajero:2932034}. The question of whether these observations represent SM physics or hint at new contributions remains an active area of investigation\cite{Schacht:2022kuj,Chala:2019fdb}.

\begin{figure}[hbtp]
\centering
 \includegraphics[width=0.6\textwidth]{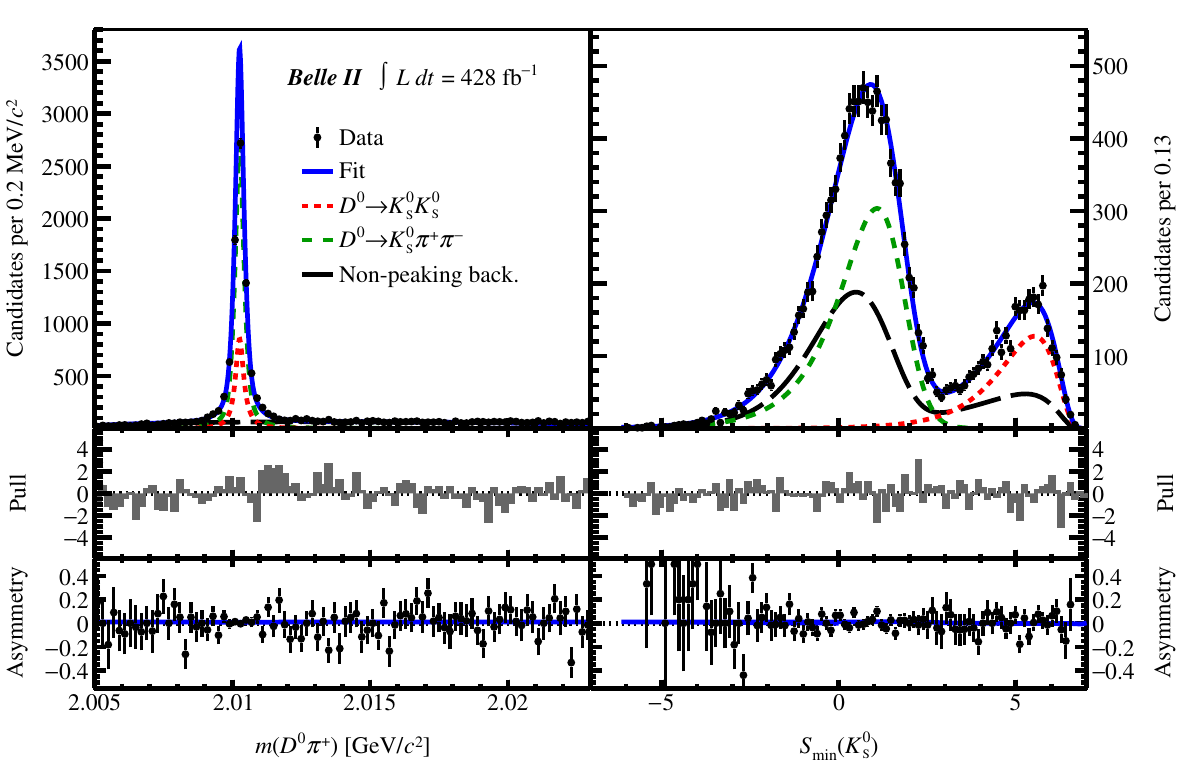}
\caption{Invariant mass distributions and resolution-weighted $K_S^0$ flight distance ($S_\text{min}(K_s^0$))  $D^0 \to K_S^0K_S^0$ decays flavour-tagged through $D^{*+} \to D^0\pi^+$ decays identified from Belle II data in Ref.~\cite{Belle:2024vho}.}
\label{fig:BellII_DstKSKS}
\end{figure}

Multiple recent measurements from both LHCb and Belle II have targeted $D^0 \to K_S^0 K_S^0$ decays. Two complementary approaches have been taken in combined analyses of  0.98~ab$^{-1}$ of Belle data and 0.43~ab$^{-1}$ of Belle II data. In the first analysis ~\cite{Belle:2024vho}, $D^0$ flavor was tagged through $D^{*+} \to D^0\pi^+$ decays, with $D^0 \to K^+K^-$ serving as a control channel for percent-level detection asymmetries. Based on a  fit to the $D\pi$ mass versus the minimum resolution-weighted $K_S^0$ flight distance, shown in Fig.~\ref{fig:BellII_DstKSKS}, yielded $A_{CP} = (-1.4 \pm 1.3 \pm 0.1)\%$. A statistically independent analysis of the same Belle and BelleII data sample employed an event-wide flavor-tagging algorithm \cite{Belle:2025cub}, fitting the $K_S^0 K_S^0$ mass versus $qr$ (where $q$ is the $D$ flavor and $r$ is the tag confidence), which gave $A_{CP} = (1.3 \pm 2.0 \pm 0.2)\%$. Combining these measurements, Belle and Belle II obtained $A_{CP} = (-0.6 \pm 1.1 \pm 0.1)\%$.

\begin{figure}[hbtp]
\centering
 \includegraphics[width=0.45\textwidth,valign=c]{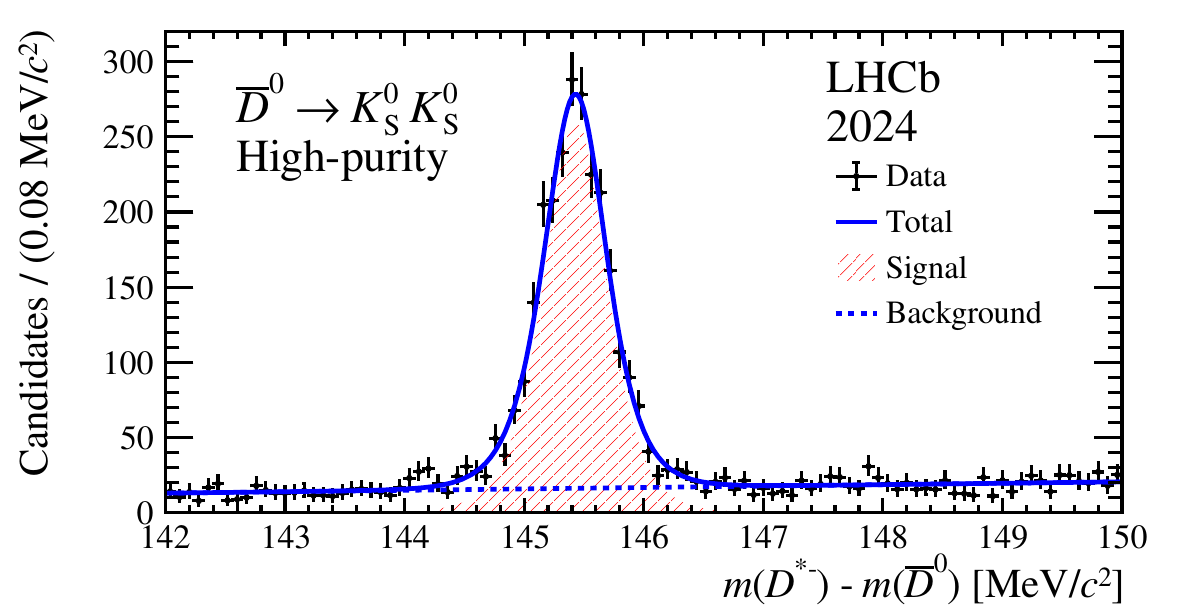}  \includegraphics[width=0.45\textwidth,valign=c]{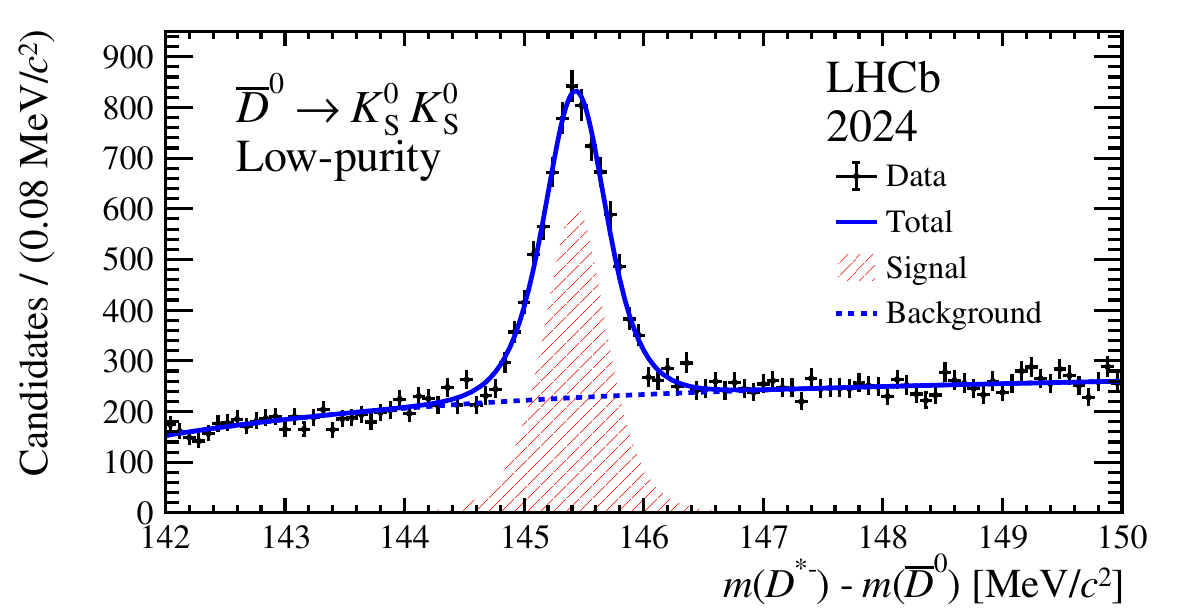}
\caption{ Projection of the three-dimensional fits to determine signal $\overline D^0\to K_S^0K_S^0$ candidates in Ref.~\cite{LHCb:2025ezf} in (left) the high-purity and (right) low-purity samples.}
\label{fig:LHCb_KSKS}
\end{figure}

LHCb reported the first result using 2024 data, employing a $6.2\invfb$ sample collected in just five months of running with the upgraded detector and software~\cite{LHCb:2025ezf}. The analysis employed $D^{*+}$ tagging and split the sample into high- and low-purity categories based on a supervised learning algorithm. Detection and production asymmetries were controlled using $D^0 \to K_S^0\pi^+\pi^-$ decays. A three-dimensional fit to $m(D^{*+}) - m(D^0)$ and the two $K_S^0$ masses yielded a signal of 15,676 $\pm$ 229 candidates. Projections of example fits are shown in Fig.~\ref{fig:LHCb_KSKS}.  The measurement determines $A_{CP} = (1.86 \pm 1.04 \pm 0.41)\%$, finding no statistically significant evidence of CP violation. The signal yield in this analysis is approximately three times as large as the yield from LHCb Run 2 analysis in the same channel \cite{LHCb:2021rdn}. This indicates a factor of three improvement in the efficiency of the Run 3 data collection with respect to the Run 2 data taking in this channel, yielding a roughly factor of 15 improvement in signal collection per unit time.

Each analysis is able to determine $A_{CP}$ in $D\to K_S^0K_S^0$ with $\mathcal O(1\%)$ precision, which will achieve sub-percent precision on this channel in a global average. This indicates fantastic prospects to probe $\mathcal O(0.1\%)$ with future datasets in this channel, probing at the same level of observed CP violation in $D^0\to h^+h^-$ decays. 

Belle II analyzed 0.43~ab$^{-1}$ of data to measure CP violation in $D^0 \to \pi^0\pi^0$ decays~\cite{Belle-II:2025rmf}. An isospin relation to the measured asymmetry in $D^0 \to \pi^+\pi^-$ decays \cite{Wang:2022nbm} implies $A_{CP} \sim \mathcal{O}(10^{-3})$ in the SM. The analysis employed $D^{*+}$ flavor tagging, with $D^0 \to K^-\pi^+$ providing control of nuisance asymmetries. Fitting the $D^0$ mass versus $m(D^{*0}) - m(D^0)$ in forward and backward regions with respect to the beam axis yielded $A_{CP} = (0.30 \pm 0.72 \pm 0.20)\%$. This result is 15\% more precise than the previous Belle measurement despite using half the integrated luminosity, demonstrating the enhanced performance of the Belle II detector and the power of the improved Belle II flavor-tagging algorithms.

\begin{figure}[hbtp]
\centering
 \includegraphics[width=0.4\textwidth]{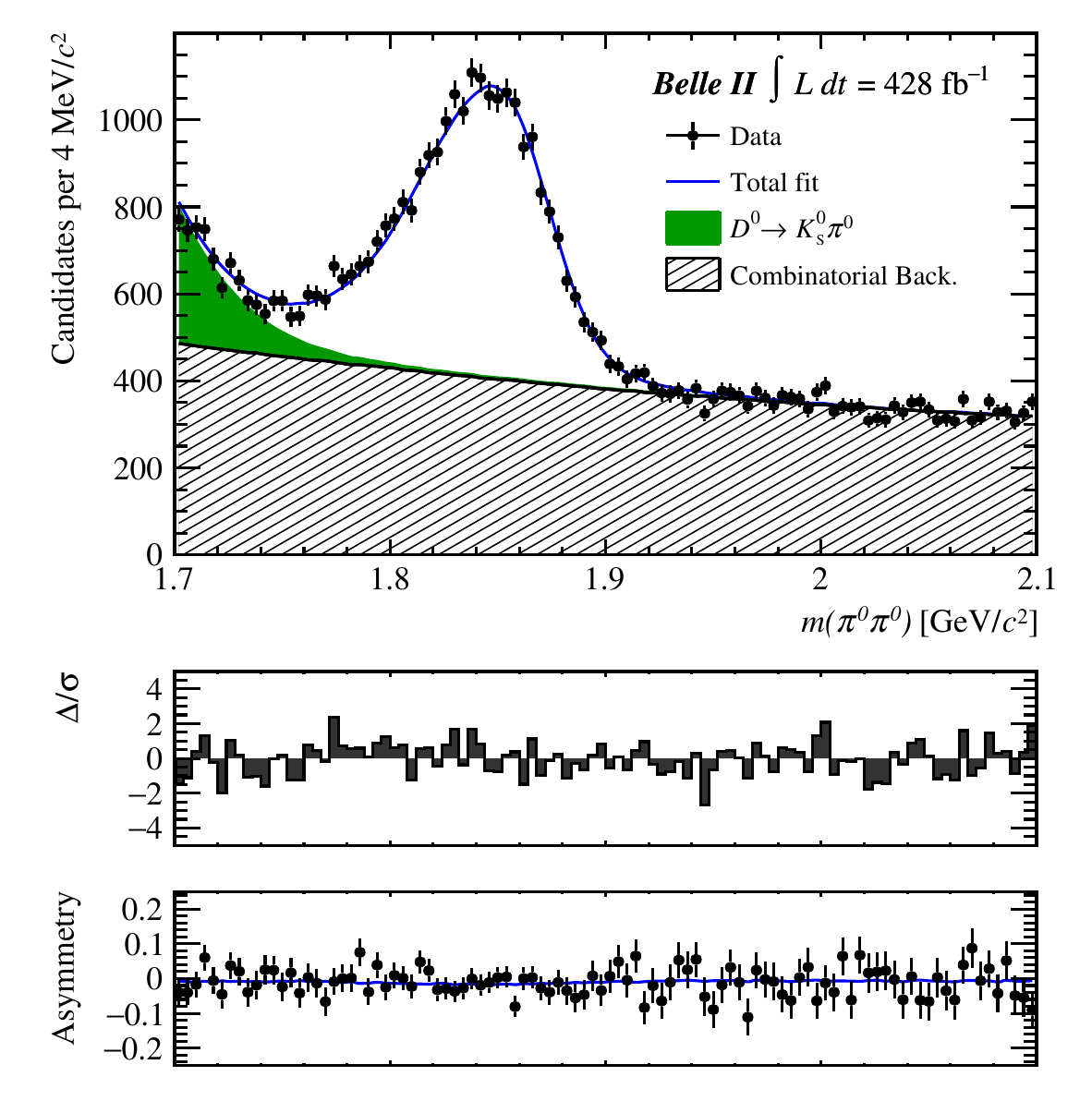}
\caption{Invariant mass distributions for (left) tagged  and (right) untagged, or``null-tag", $D^+ \to \pi^+\pi^0$ decays from Belle II in Ref.~\cite{Belle-II:2025rmf}.}
\label{fig:d_pi0pi0}
\end{figure}

Belle II also examined the $D^+ \to \pi^+\pi^0$ decay with 0.43~ab$^{-1}$ sample, split between $D^{*+} \to \pi^0 D^+$-tagged and untagged $D^+$ candidates~\cite{Belle-II:2025wsy}. Due to the fact that only a single weak amplitude contributes to this decay, the SM predicts no CP asymmetry in this channel, and thus an observation would constitute unambiguous evidence for physics beyond the SM. The measurement obtained $A_{CP} = (-1.8 \pm 0.9 \pm 0.1)\%$, providing the most precise measurement to date and in agreement with previous measurements from Belle \cite{Belle:2017tho} ($A_{CP} = 2.3 \pm 1.2 \pm 0.2\%$) and LHCb Run 1+2 \cite{LHCb:2021rou} ($A_{CP}  =-1.3 \pm 0.9 \pm 0.6\%$).

\begin{figure}[h]
\centering
 \includegraphics[width=0.6\textwidth]{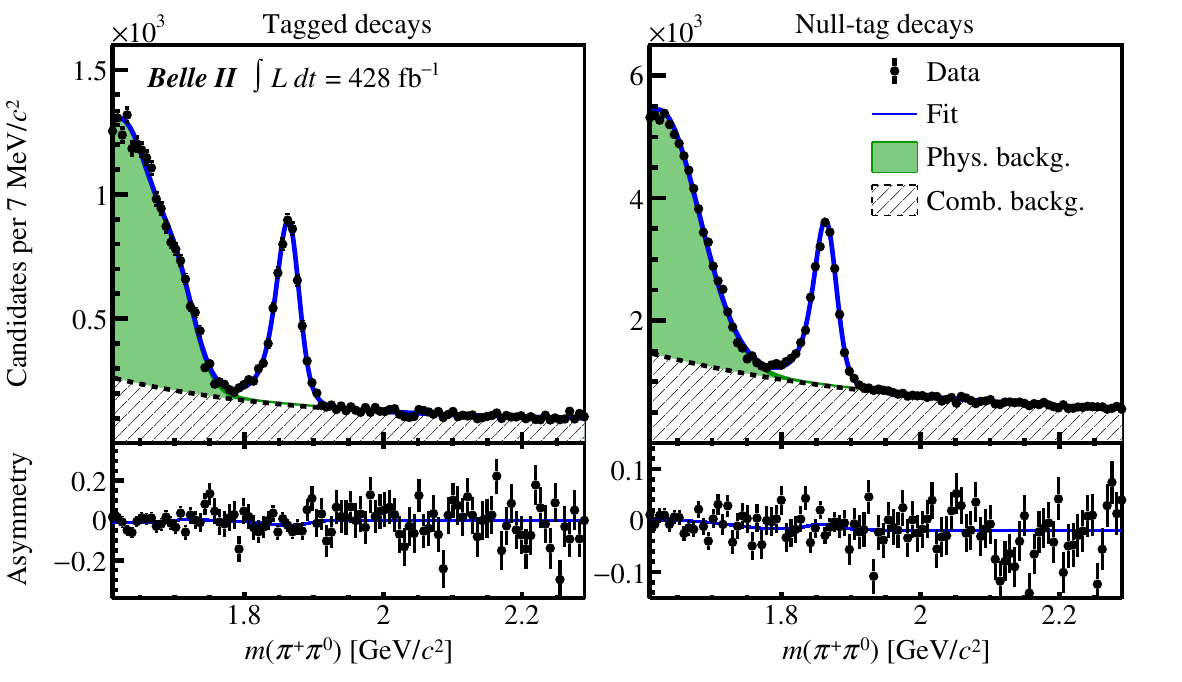}
\caption{Invariant mass distributions for (left) tagged  and (right) untagged, or``null-tag", $D^+ \to \pi^+\pi^0$ decays from Belle II from Ref.~\cite{Belle-II:2025wsy}.}
\label{fig:dplus_pipi0}
\end{figure}

\section{Conclusions}

Time-integrated measurements of CP violation provide rigorous tests of the SM paradigm of CP violation arising from a single complex phase in the CKM matrix. Recent progress has been substantial across multiple fronts.

Measurements of the CKM angle $\gamma$ through $B \to Dh$ decays have achieved a global precision of approximately $2.5^\circ$, with the combined result $\gamma = (65.7 \pm 2.5)^\circ$ showing excellent consistency across numerous complementary measurements. 

The observation of CP violation in baryon decays represents a major milestone. The absence of large effects in two-body $\Lambda_b^0 \to ph^-$ decays, contrasted with the significant asymmetries observed in resonant structures of $\Lambda_b^0 \to \Lambda K^+K^-$ and $\Lambda_b^0 \to pK^-\pi^+\pi^-$, demonstrates that strong dynamics play a crucial role in manifesting CP violation in baryon systems. Theoretical progress in understanding these strong interaction effects is essential for fully exploiting the discovery potential of baryonic CP violation.

In the charm sector, measurements approach sub-percent precision on key observables. The global averages for $A_{CP}$ in $D^0 \to K_S^0 K_S^0$, $D^0 \to \pi^0 \pi^0$, $D^+ \to \pi^+ \pi^0$ decays are expected to achieve sub-degree precision with available results. The puzzles in $D^0 \to h^+h^-$ asymmetries continue to motivate further experimental and theoretical investigations.

Looking forward, substantial increases in data are expected from both LHCb (targeting 23~fb$^{-1}$ by the end of 2026) and Run 2 of Belle II. LHCb's Run 3 performance, exemplified by the factor-of-15 improvement in signal-per-unit-time  $D^0 \to K_S^0 K_S^0$ events relative to Run 1 and Run 2, demonstrates the dramatic gains from the upgraded detector and trigger system. The CMS and ATLAS experiments also hold exciting prospects for measurements of CP violation in data collected in the current Run of the LHC.

The combination of increased statistics, refined analysis techniques, and complementary measurements from multiple experiments positions the field to probe CP violation with unprecedented precision with datasets that are currently being collected. The results that are achieved could reveal subtle deviations from SM predictions that indicate Beyond-Standard-Model physics.

\bibliographystyle{LHCb}
\bibliography{main}

\end{document}

%% file: definitions.tex
\usepackage{xspace} 

\def\beq{\begin{equation}}
\def\eeq{\end{equation}}

\def\DSTO{\ensuremath{D^{*0}}\xspace}

\def\aDSTO{\offsetoverline{D}^{*0}}

\def\invfb{\text{ fb}^{-1}}

\def\thebaroffset{0.1em}
\newcommand{\offsetoverline}[2][\thebaroffset]{\kern #1\overline{\kern -#1 #2}}%

\def\D       {{\ensuremath{D}}\xspace}

\def\Dp      {{\ensuremath{\D^+}}\xspace}
\def\Dm      {{\ensuremath{\D^-}}\xspace}

\def\DpDm    {\ensuremath{\Dp {\kern -0.16em \Dm}}\xspace}

\def\Dstarz  {\DSTO\xspace}
\def\Dstarzb {\aDSTO\xspace}

\def\DstarzDstarzb    {\ensuremath{\Dstarz {\kern -0.07em \Dstarzb}}\xspace}

\newcommand{\aunit}[1]{\ensuremath{\text{\,#1}}}       
    
\def\fb   {\ensuremath{\aunit{fb}}\xspace}
\def\invfb   {\ensuremath{\fb^{-1}}\xspace}
 
